\documentclass[runningheads]{llncs}

\usepackage{graphicx}
\usepackage{booktabs}
\PassOptionsToPackage{hyphens}{url}\usepackage{hyperref}
\usepackage{multirow}
\usepackage{subcaption}
\usepackage{csquotes}
\usepackage{enumitem}
\newlist{RQ}{enumerate}{1}
\setlist[RQ]{label=RQ\,\arabic*: ,ref={RQ\,\arabic*: }}
\begin{document}
\title{Training Software Engineers for Qualitative Evaluation of Software Architecture}
\titlerunning{Evaluating software architecture from a Quality perspective}
\author{Ritu Kapur\inst{1}
%\orcidID{0000-0001-7112-0630} \and
Sumit Kalra\inst{2} \and
Kamlesh Tiwari\inst{3} \and
Geetika Arora\inst{3}}
\authorrunning{R. Kapur et al.}
\institute{Indian Institute of Technology Ropar, Punjab, India. 
\email{dev.ritu.kapur@gmail.com}\\
%\url{https://sites.google.com/view/ritu-kapur} 
\and
Indian Institute of Technology Jodhpur, Rajasthan, India.
\email{sumitk@iitj.ac.in}\\
\and
Birla Institute of Technology and Science Pilani, Rajasthan, India.\\
\email{\{kamlesh.tiwari,p2016406\}@pilani.bits-pilani.ac.in}}
%
%\author{Anonymous}
%\authorrunning{Anon. et al.}
\maketitle              % typeset the header of the contribution
\begin{abstract}

 A software architect uses quality requirements to design the architecture of a system. However, it is essential to ensure that the system's final architectural design achieves the standard quality requirements. The existing architectural evaluation frameworks require basic skills and experience for practical usage, which the novice software architects lack.
 
 We propose a framework that enables novice software architects to infer the system's quality requirements and tactics using the software architectural block-line diagram. The framework does not assume any specific type of architectural diagrams. It takes an image as input, extracts various components and connections, and maps them to viable architectural patterns, followed by identifying the system's corresponding quality attributes (QAs) and tactics. The framework includes a specifically trained machine learning model based on image processing and semantic similarity methods to assist software architects in evaluating a given design by a) evaluating an input architectural design based on the architectural patterns present in it, b) lists out the strengths and weaknesses of the design in terms of QAs, c) recommends the necessary architectural tactics that can be embedded in the design to achieve the lacking QAs. 
 
 To train our framework, we developed a dataset of 2,035 architectural images from fourteen architectural patterns such as Client-Server, Microservices, and Model View Controller,  available at \url{https://www.doi.org/10.6084/m9.figshare.14156408}.
 The framework achieves a Correct Recognition Rate of 98.71\% in identifying the architectural patterns with fourteen classes. We evaluated the proposed framework's effectiveness and usefulness by using controlled and experimental groups, in which the experimental group performed approximately 150\% better than the controlled group. The experiments were performed as a part of the Masters of Computer Science course in an Engineering Institution.

\end{abstract}
\section{Introduction}
\label{sec:introduction}
Software development generally begins with the specification of the functional and non-functional requirements of the software. Both the functional and non-functional requirements are listed in the requirements specification of software. The inter-dependency of the functional and non-functional requirements makes both critical for software quality. The non-functional requirements are specified as quality attributes (QAs), such as performance, security, and reliability. The functional requirements model the functional correctness of a software, while the non-functional requirements capture the degree to which a software achieves the intended functionality \cite{chung2012non}. 

QAs have been termed as reusable architectural building blocks used in developing an application architecture \cite{kim2009quality}. However, in order to embed the QAs in software design, one implements various software tactics. For instance, \emph{authentication} tactic is incorporated into a software design to attain \emph{security}; \emph{queues} are incorporated to attain better \emph{performance}, and so on. The design choices made by software developers among various architectural patterns and their tactics determine the software quality. %Therefore, it becomes essential to evaluate the software architecture by analyzing the architectural patterns and tactics it comprises and the QAs it fulfills. 

Novice software architects often lack the knowledge and expertise of using various software tactics to embed different software patterns in an architecture. Further, traditional teaching methods do not provide much opportunity to teach the design essentials and their effective implementation to the novice software architects \cite{capilla2020teaching,pinto2017training}. Thus, it becomes challenging to make adequate design decisions in selecting the appropriate design alternatives while developing a software. However, while performing a software architecture evaluation (SoftArchEval), it becomes essential to validate if a particular software design fulfills the required QAs and meet the necessary quality standards.  

\subsection{Motivation}

Broadly, the prime objective of our work is to develop a framework that:
\begin{enumerate}
    \item Given an input architectural design image, determines and lists the QAs it meets or lacks, and
    \item Recommends the relevant software tactics to achieve the lacking QAs.
\end{enumerate}

%Software architecture primarily comprises a combination of several architectural patterns, each bearing its peculiar characteristics. 
For instance, the pipe and filter architectural pattern distributes a task's processing into various independent sub-tasks and is thus primarily used to attain performance, availability, and reliability.  Timely detection of software quality by analyzing its architectural design can lower the overall defect-fixing cost \cite{kapur2020defect}. Some of the critical questions in this context could be:

\begin{enumerate}
 %\begin{RQ}[align=parleft, leftmargin=!,itemsep=1pt,labelsep=12pt]
    \item What are the significant QAs and tactics associated with architectural design?
    \item How to evaluate the architectural designs using QAs? 
    \item What are the necessary architectural tactics to achieve various QAs?
    \item Is there any automated method to detect the standard architectural patterns present in an architectural design?
    \item Given an input architectural design image, is it possible to detect the QAs it captures or lacks in and the necessary tactics that can be implemented to achieve the lacking QAs? 
%\end{RQ}
\end{enumerate}

This paper addresses these questions via an in-depth examination of QAs associated with various architectural design patterns and the necessary tactics to implement them.

\subsection{Broad Idea of our work}
\begin{figure}
\centerline{\includegraphics[scale=0.8]{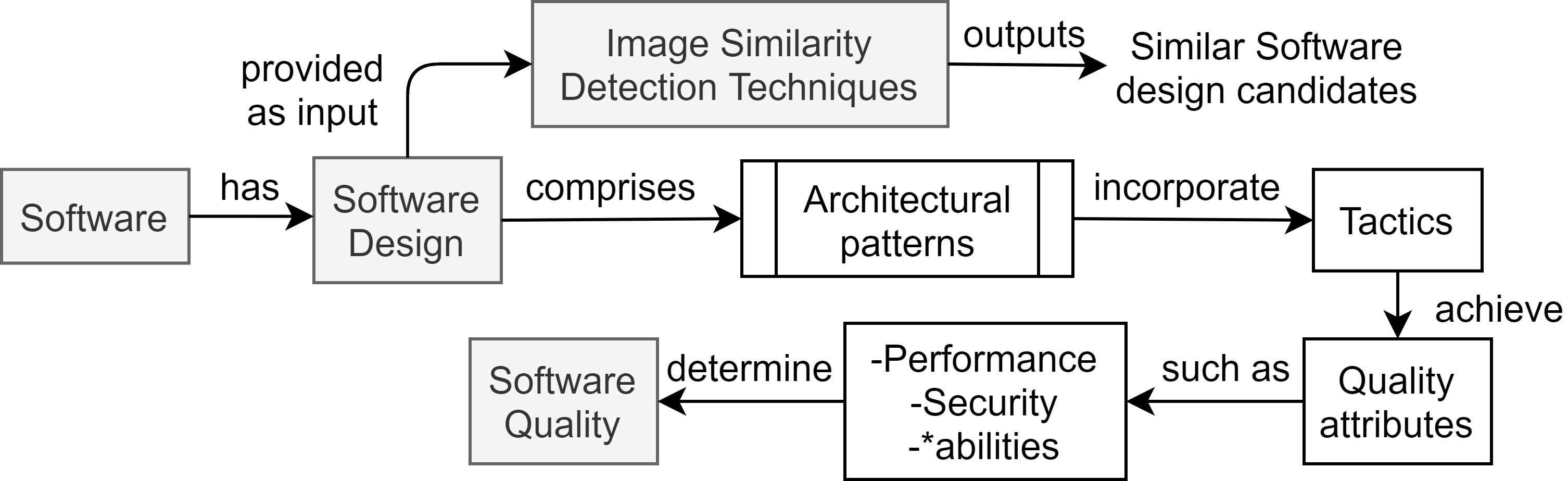}}
\caption{Broad idea of the proposed approach}
\vspace{-0.56cm}
\label{fig:basic-idea}
\end{figure}

Fig. \ref{fig:basic-idea} illustrates the central idea underlying our approach:

\begin{enumerate}
    \item Every software has an architectural design, which comprises one or more patterns.
    
    \item Similar software designs can be identified using the existing Image Similarity detection techniques.  
    
    \item Every architectural pattern may  incorporates several software tactics. For instance, the \emph{pipe-and-filter pattern} incorporates \emph{fault detection,} \emph{recovery,} and \emph{performance} tactics by various design concepts, such as ping (or echo), active redundancy, shadowing, spare, and so on.  
    \item Successful execution of software tactics leads to the fulfillment of QAs. For instance, the successful execution of recovery and performance tactics leads to the \emph{availability,} \emph{reliability,} and \emph{maintainability} of QAs. 
    
    \item  Software quality is defined as the degree to which a system, component, or process meets the specified requirements \cite{ieee1990ieee}.
    \item Software QAs metrics can be used to measure software quality.
\end{enumerate}
\pagebreak
\section{Related Work}
\label{sec:related-work}
%Section \ref{sec:existing-soft-eval-methods} provides a categorization of the existing evaluation methods. In this section, we discuss some of the existing evaluation techniques related to our work.

\subsection{Broad Categorization of SoftArchEval methods}
\label{sec:existing-soft-eval-methods}
A broad categorization of the existing evaluation methods \cite{abowd1997recommended,dobrica2002survey} is as follows:

\begin{enumerate}
    \item \emph{Questioning methods} comprise asking qualitative questions on an architecture derived from the software quality aspect. These can be used to evaluate the quality of any given architecture and are classified as follows:  
    \begin{enumerate}
        \item \emph{Scenario: } It is a sequence of steps involving the use or modification of the system. It provides a means to characterize how well a particular architecture responds to the demands placed on it by those scenarios. 
        
        \item \emph{Questionnaire:} It is a list of general and relatively open questions applicable to all architectures. The questions generally cover various topics, such as the architectural generation method or the architectural description details. 
        
        \item \emph{Checklist:} It is a detailed list of questions developed after evaluating a domain-specific set of systems.
    \end{enumerate}
    \item \emph{Measuring methods} comprise quantitative measurements made on the architecture by addressing specific QAs, and have the following salient types: 
     \begin{enumerate}
         \item  \emph{Experience-based methods:} These are based on the experiential knowledge of the experts  \cite{bosch2000design}. 
          \item  \emph{Metrics-based methods:} Metrics are termed as the quantitative interpretations of some observable measurements on architectural elements, such as the fan-in/ fan-out of various components. The metrics-based evaluations focus on the metrics values extracted from various architectural components while considering the assumptions involved in such metrics \cite{selby1995interconnectivity}. 
           \item \emph{Mathematical modeling-based methods:} These comprise mathematical proofs and methods for evaluating the operational QAs, such as the performance and reliability of architectural components \cite{reussner2003reliability}.
           \item \emph{Simulations, prototypes, and experiment-based methods:} Simulations, prototypes, and experimental results are often a part of the software development process and play an important role in answering various questions during software reviews. For instance, simulation results can be used to validate specific assertions. However, this tends to be an expensive approach if a prototype is specifically developed to perform the evaluation \cite{bengtsson1998scenario}.
     \end{enumerate}
    
\end{enumerate}

\subsection{Generic SoftArchEval works}
The \emph{scenario-based Architectural Analysis method  (SAAM)} \cite{kazman1994saam} was introduced in 1993 to describe and analyze the software architecture based on various QAs. It was stated that software architectural analysis could help detect software defects in the early phases of software development, reducing the overall cost. With the emphasis on different QAs, different SAAM versions were developed. For instance, SAAM majorly focused on modifiability, while \emph{SAAM-founded on Complex Scenarios (SAAMCS)} \cite{lassing1999software} which was an extension of SAAM emphasized on flexibility. Similarly, \emph{Extending-SAAM by Integration in the domain (ESAAMI)} \cite{molter1999integrating}  is an improved version of SAAM, which combines analytical and reuse concepts and integrates the SAAM in the domain-specific and reuse-based development process, and \emph{Software Architectural Analysis Method for Evolution and Re-usability (SAAMER)} \cite{lung1997approach} is an extension of SAAM emphasizing \emph{evolution} and \emph{reusability}.  %The method performed the software analysis using three different dimensions, viz., the software's functionality, the structure or design of the software, and the allocation choices made to embed various functionalities in the system. The analysis assesses the architectural assumptions made while making various design choices in developing the software and the inherent risks involved. SAAM evaluates software by analyzing it in various scenarios and determining if a particular scenario requires any architectural changes to the software. The scenario development is performed by various stakeholders, who identify the viable scenarios. The scenarios that require architectural changes are indirect, and those that do not require any changes are called direct. SAAM expresses various scenarios based on the modifications required and associates them with the cost of respective modifications. The costs serve as the basis of the overall evaluation of the candidate architectures. SAAM has been validated in various case studies, such as air traffic control, user interface development environments, Internet information systems, keyword in context (KWIC) systems, and embedded audio systems.

\emph{The Architectural Trade-off Analysis Method (ATAM)}\cite{kazman1998architecture} provides a framework for evaluating software architecture concerning multiple QAs, particularly modifiability, performance, availability, and security. ATAM introduces the notion of tradeoff among multiple QAs, given a software architectural description. ATAM requires a software architectural description based on Kruchten's \enquote{$4+1$} views \cite{kruchten19954+} and requires several views, viz., a dynamic view, a system view, and a source view. %It also considers various scenarios, such as use cases, growth scenarios, and exploratory scenarios, and qualitative analysis heuristics to perform the evaluation. The evaluation process comprises various scenarios and requirements, architectural views and scenario realization, attribute model building and analysis, and tradeoffs. The attribute analysis leads to the discovery of various tradeoff points and sensitivity points. A tradeoff point is a property that affects more than one attribute and is a sensitivity point for at least one attribute. ATAM provides for an iterative improvement strategy. At the end of the evaluation process, the analysis results are compared with the system requirements. If the system-predicted behavior is reasonably close to the requirements, the designers are signaled to proceed with the detailed design phase; else, they are notified to develop an action plan for changing the architecture. ATAM has been applied to several software systems and is under research.
\emph{Quality-Driven Architecture Derivation and Improvement (QuaDAI)} \cite{gonzalez2013defining}  is a metrics-based method used for derivation, evaluation, and improvement of software product architectures obtained in Software Product Line (SPL) development processes. QuaDAI performs the evaluations based on a) SPL viewpoints: functional, variability, quality, transformation, and b) and a process consisting of a set of activities conducted by model transformations to allow the automatic derivation, evaluation, and improvement of a product architecture from the SPL architecture. %The procedure comprises:
%\begin{enumerate}
%\item deriving the product architecture from SPL architecture to maximize the QA-requirements,
%\item evaluating the product architecture by measuring the QA fulfillment as per the requirements, and
%\item transforming the product architecture to meet the QA-requirements.
%\end{enumerate}
%The authors report that QuaDAI achieves better results than ATAM when evaluated by novice software architectural evaluators, and there exists a potential to improve the method's usability.

\subsection{Training novice software architects}
Some recent works have felt the need of devising better methods for training software engineers to improve their decision making process \cite{pinto2019training}. Automated solutions for software design and development can help improve the decision making process of novice software architects or undergrad students \cite{kapur2019towards,7332519,kapur2020defect,kapur2021using}. However, the existing studies have developed solutions to facilitate the collaborative decision making process and study the effect on cognitive and modelling tasks \cite{capilla2020teaching}. In some of the existing studies proposing software solutions for training students, it has been reported to result in a positive boost of student's confidence and the improvement of their technical skills\cite{pinto2017training}.

\subsection{Recovering Architectural information from design diagrams}
The Image Extractor for Architectural Views (IMEAV) \cite{maggiori2014towards} is proposed to extract the architectural views from design images. However IMEAV is only applicable to Unified Modeling Language (UML) design diagrams, and cannot be used to detect architectural patterns in other type of design images. Similarly, most of the other existing studies are focused on the detection of images based on photographic scenes (based on image depths) \cite{jin2021image}, Synthetic Aperture Radar (SAR) images (based on Intensities) \cite{ye2017robust}, or on high-speed tracking by detection \cite{bochinski2017high}. However, the applications of these works and the different image types considered differ considerably from architectural design diagrams. However, it has been validated by some of the existing studies that SIFT when tuned effectively, acts as an effective image detection technique \cite{jin2021image}.

\textbf{Limitations of the existing works:}
\begin{enumerate}
    \item Most of the existing SoftArchEval works consider a subset of QAs for software evaluations. 
    \item The complexity in the existing SoftArchEval methods limit their use. Also, there is an overhead of learning the evaluation methods and training the novice software architect to use them.
    \item Almost all existing methods require inputs from stakeholders, software architects, and various experts. The dependency comes with an additional cost and limits the scope of knowledge to the involved participants' experience.  
    \item Most of the image datasets are based on photographic scenes or non-architectural design images.
    \item To the best of our knowledge, none of the existing works recommend the necessary software tactics to achieve various QAs. Knowledge of the essential software tactics used for specific QAs is unexplored in the existing works.
\end{enumerate}

To overcome the limitations listed above, we propose a machine-learning-based assisting framework for software architectural evaluation that leverages the knowledge present in architectural images and the relationship between various architectural patterns and QAs to perform the evaluations. Using the relationship between various software tactics and QAs, our framework also recommends the necessary tactics to achieve specific QAs. We have not come across any work exploring such research direction for architectural evaluation to the best of our knowledge.   

\begin{figure}
\centerline{\includegraphics[scale=0.75]{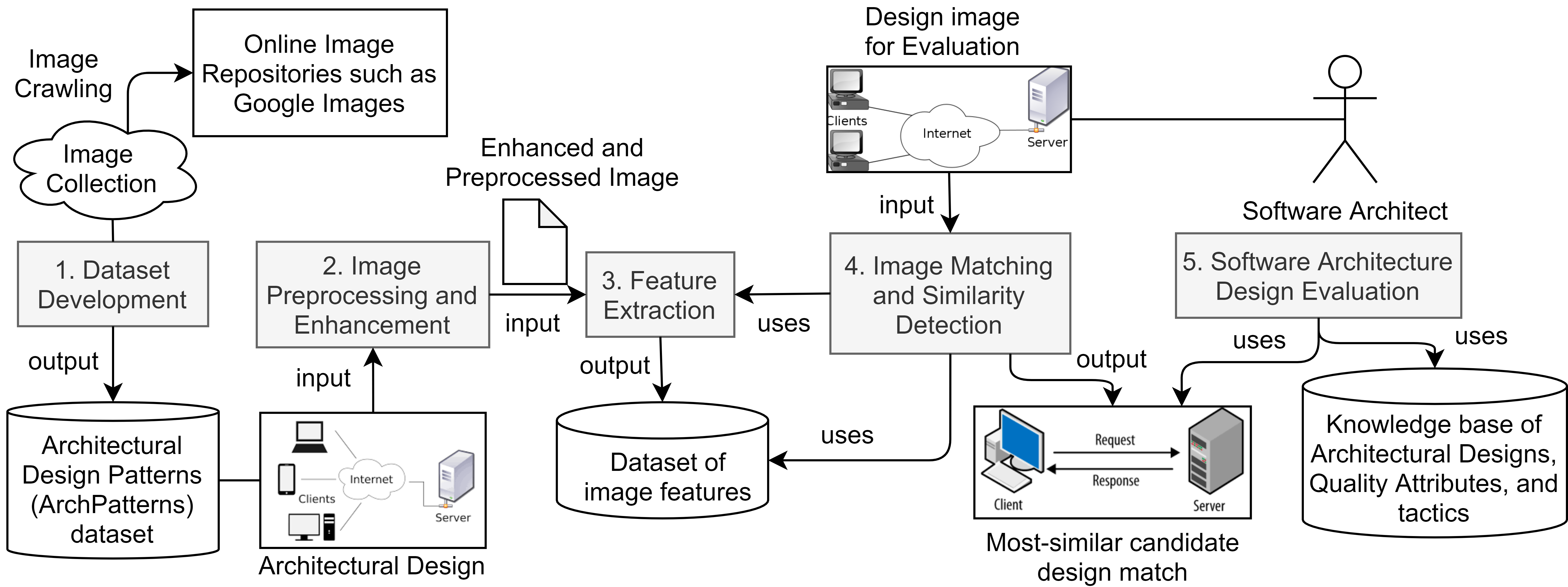}}
\caption{An overview of the proposed method}
\vspace{-0.57cm}
\label{fig:framework}
\end{figure}

\pagebreak
\section{Proposed Methodology}
\label{sec:proposed-method}
Fig. \ref{fig:framework} illustrates the essential stages involved in implementing our approach:

\begin{enumerate}
    \item \textbf{Dataset development:} \label{step:image-dataset} We developed an Architectural Design Patterns dataset, named as ArchPatterns dataset, comprising 2,035 architectural design images from fourteen different architectural patterns viz., broker, layered, event-bus, pipe-and-filter, repository, microkernel, microservices, model-view controller, peer-to-peer, presentation abstraction controller, client-server, space-based, representational state transfer (REST), and publisher-subscriber. We considered more than 100 design images in each pattern category. 
    The images were collected from online sources such as official blogs and technical write-ups with ground truth. After collecting these images, we manually filtered them to remove the images not projecting the relevant design patterns or not of the desired quality. Ground truth for the remaining images is manually annotated for the training. Images are of three channels with RGB color coding of different resolutions and sizes.
     We consider the QAs and sub-QAs listed by ISO/ IEC 25010 standard as a reference point for our study. The necessary tactics required to achieve these QAs were derived by reviewing some standard software architectural reference books  \cite{bass2003software,bachmann2003deriving,scott2009realizing,kalra2018towards,bachmann2003deriving,kim2009quality,li2020understanding,osses2018exploration,bi2018architecture,mark2015software,trowbridge2004integration,fielding2000architectural,bachmann2007modifiability} and the information is stored in the form of structured tables available at \url{https://doi.org/10.6084/m9.figshare.14623005}.
     Our ArchPatterns dataset is publicly shared at \url{https://doi.org/10.6084/m9.figshare.14156408}.

    \item \textbf{Image Preprocessing and Enhancement:} To improve the quality of our images and remove noise, we preprocessed all the ArchPatterns dataset images and scaled them to  same size. We considered the images of only a considerable resolution ($>=350*350$ pixels) to have an effective image detection and matching. 
    
    \item \textbf{Feature Extraction:} \label{step:feature-extraction} We extract the image features, software components, their interconnections, and component labels present in the ArchPatterns dataset. To speed-up the image matching (or lookup) task during the evaluation stage, we also store the similarity of an image when compared to all the rest present in the dataset. We store these image features as a structured dataset, named as \emph{ImageFeatures} dataset. %Feature extraction ensures that only the images' salient features are considered for matching instead of comparing the whole image. 
    We used SIFT \cite{lowe1999object}, a scale-invariant interest point detector with corner properties at different scales for feature extraction. SIFT also devises a suitable feature descriptor to uniquely represent the interest points. 
    \item \textbf{Image matching and Similarity detection:} \label{step:image-matching} Every image has multiple and different numbers of interest points. Interest point descriptors across different architectural images are subjected to achieve a mutual similarity, and then the count of matching interest points is used as a similarity measure between the two architectural images. For instance, if the images A and B having $N_a$ and $N_b$ number of SIFT interest points have $N_{ab}$ number of highly correlated descriptors in common, then the match score is computed as:
    \begin{equation}
 \label{eq:1}
      Score = 1 - \frac{N_{ab}}{min(N_a , N_b)}
 \end{equation}
    Each image in the database is matched with all other database images to obtain a dis-similarity score during testing. The images belonging to the same architectural pattern are expected to have a low score, whereas those belonging to different pattern classes have a high score value. The matching score of every pair of images is obtained, and the score list is sorted in increasing order of dis-similarity. Further, for a given query architectural image, the most similar architectural image in the database is determined, and the corresponding label is assigned to the query image. The suitability of SIFT lies in the fact that the method builds a scale-space pyramid and only chooses prominent feature points that appear in all scales. By this, it achieves scale-invariance, which is a much-needed property for the architecture images. The same is evident from the result of achieving a higher correct recognition rate (CRR), which is defined as the percentage of images, out of total images, for which the recognition is correct at Rank-1 retrieval. Suppose out of $n$ test images, $x$ images are found to be true matches at Rank-1, then:

%\noindent    

    \begin{equation}
    \label{eq:CRR}
      CRR= (x/n)\times 100
    \end{equation}

 \item \textbf{Software Architecture Design Evaluation: } \label{step:eval} When a software architect starts evaluating an architectural design image using our framework, the steps involved are:
 \begin{enumerate}
     \item \emph{Feature extraction} of the image as described in Step \ref{step:feature-extraction}.
     \item Using the extracted features, the \emph{image matching and similarity detection} is performed to determine the top-similar match from the  \emph{ImageFeatures} dataset. This step provides the necessary information about the most-likely architectural pattern prominent in the considered design image. 
     \item \emph{Architectural Evaluation:} The software architect can then conclude about the strengths and weaknesses of the design (in terms of QAs) using our knowledge-based provided in the form of tables (discussed in the Section \ref{sec:proposed-method}).  The software architect can then work in the direction of improving the design by implementing the necessary tactics as listed by the tables.  
 \end{enumerate}

\end{enumerate}

 \section{Experimental System}
   The essential objective of our work lies in improving the decision making process of software architects by improving the understanding behind software architectural patterns, QAs, and tactics. We achieve our goal by providing a framework for supporting the architectural design decision making process where:
   \begin{itemize}
       \item The framework's \emph{image detection and matching module} helps the novice software architects in determining the software architectural patterns in a design, and 
       \item The framework's \emph{knowledge-base} present in the form of tables provides the necessary information of the software tactics required to embed various QAs in the design.
   \end{itemize}
   
   The research questions that guide this study are:
   \begin{enumerate}
       \item What is our framework's highest CRR in identifying the software architectural design patterns?
       \item How does our framework impact the decision making process of the software architects?
   \end{enumerate}
   \vspace*{-1cm}
  % Effective image detection and matching plays a major role in our framework. Hence, one of our 
    
      \begin{figure}[hbtp]
        \centering
        \includegraphics[scale=0.4]{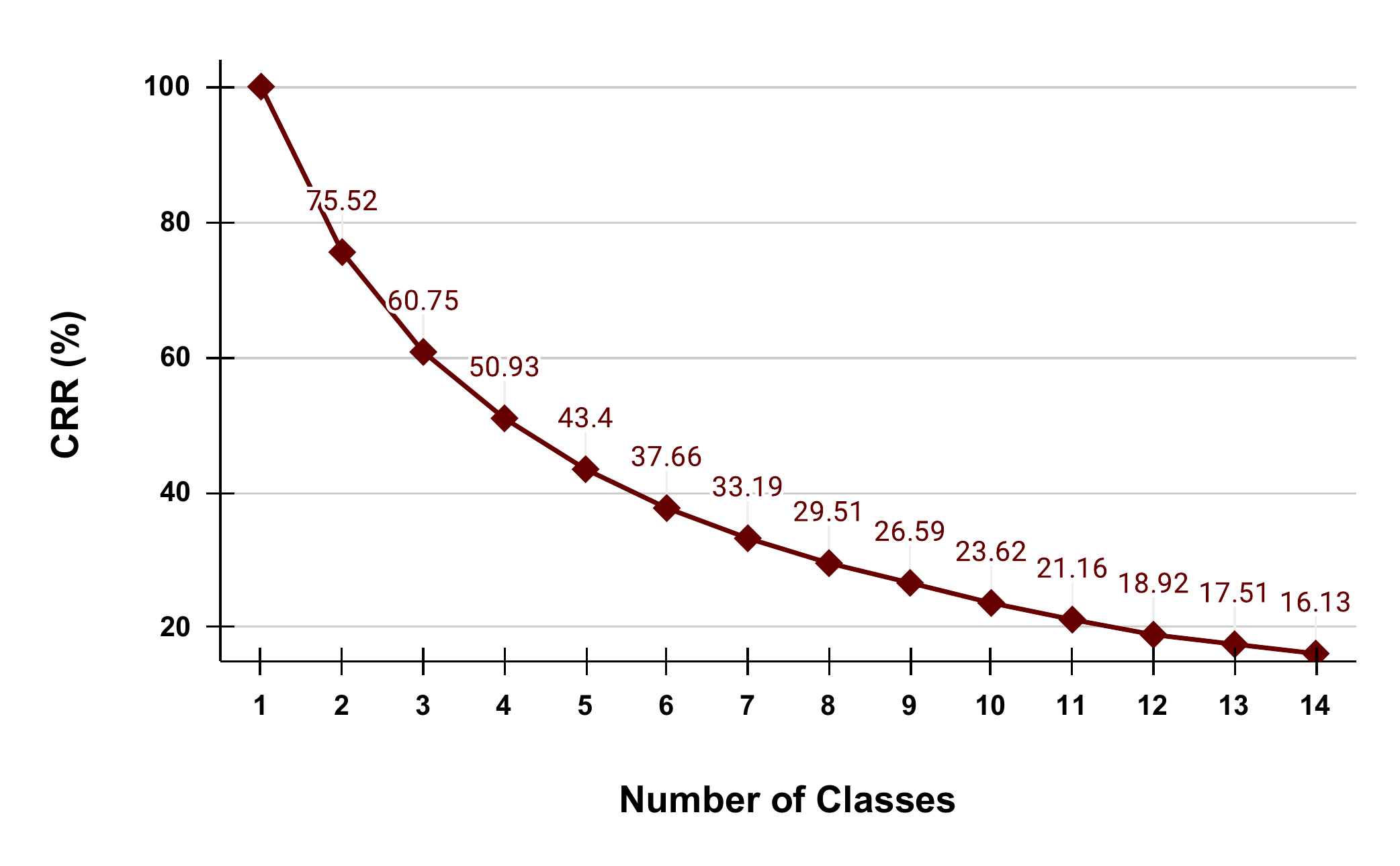}
        \vspace*{-0.5cm}
        \caption{Effect of the number of classes on the CRR}
        \vspace*{-0.5cm}
        \label{fig:result_matching}
    \end{figure}
    
    \subsection{Experimental setting and results analysis for image matching}  
     Architectural images in our database are scaled to the size $100\times 100$. We followed the one-vs-rest matching strategy where each image in the database is matched with all the remaining images to obtain the matching score. Since we have 2,035 images in the database, we obtained 4133088 scores that contain 337958 genuine and 3795130 imposter matching scores. Our framework achieves a Correct Recognition Rate (CRR) of 17.51\% in identifying the architectural patterns when all the classes have been considered. However, CRR keeps increasing when fewer classes are considered and 75.52\% when architectural images of any two classes are compared. The highest CRR obtained with our framework is 98.71\%. The change in CRR value with a change in the number of classes is annotated and depicted using the graph shown in the Fig. \ref{fig:result_matching}. Here, the classes represent the architectural patterns (discussed in Section \ref{sec:proposed-method}) considered in this work.

    \subsection{Experimental setting for the experiments with software architects}
    
    As part of a post-graduate course on Software and Data Engineering, we identified 64 suitable participants for this study such that 30 participants were selected who were having IT industry exposure of more than one year and rest of the 34 participants were not having any industry exposure or less than one year experience. All the participants were taught IEC/ISO 25010 Software Quality Models and Design Patterns through online class lectures. To conduct the study, we divided each of this group in equal halves and formed controlled and experimental groups as shown in Table \ref{tab:groups}.
    During the experiment, all the participants were provided the access to course lectures and teaching material on ISO 25010 and Design Patterns. However, for the experimental group, an additional access to the image-processing tool interface and the knowledge base was also provided. This was done to study the impact on decision making process due to our framework's support. 
   \vspace*{-1cm}
    \begin{table*}
    \caption{Groups and participants}
    \label{tab:groups}
\begin{center}
\resizebox{0.7\textwidth}{!}{
\begin{tabular}{c|c|c}
\toprule
\multirow{2}{*}{\textbf{\shortstack{Participant Type\\ or Group}}}&\multicolumn{2}{c}{\textbf{Type of Experimental Setting}}\\ 
\cline{2-3}
 & \textbf{Experimental}& \textbf{Controlled}\\
 \midrule
 \textbf{Without Industry Experience} & 17 & 17\\ \hline
 \textbf{With Industry Experience} & 15 & 15\\
 \bottomrule
\end{tabular}}
\end{center}
\end{table*}
   \vspace*{-1cm}

\subsection{Evaluation Metrics}
The proposed system provides QA analysis for a given architectural diagram of a software system. To evaluate the proposed framework's effectiveness, we conducted experiments by sharing the generated QA analysis with experimental group participants and compared the performance with control group participants. As part of the experiment, we ask the following questions:
\begin{enumerate}
    \item \emph{Accuracy:} How accurately experimental group participants performed in terms of QA analysis's correctness compared to the control group participants?  
    \item \emph{Time Performance:} Is there any improvement in the time taken to analyze a given software architectural diagram by the experimental group participants compared to control group participants? 
    \item \emph{Explainability:} How much appropriate reasoning/ explanation was provided for the experimental group's analysis compared to the control group participants? 
    \item \emph{Robustness: } How many different  architectural patterns can be processed more accurately with the help of the proposed system by the experimental group compared to the control group? We selected architectural diagrams from 14  categories of architectural patterns. Our framework is extendable and can consider more categories given the corresponding data for architectural diagrams and descriptions.  
\end{enumerate}

%\begin{figure*}
%\centering
%\hspace*{-6em}
%\begin{subfigure}{0.5\textwidth}
%\rule{\linewidth}{0.5\linewidth}
 % \centering
  %\includegraphics[width=1.1\linewidth]{figures/accuracy.pdf}
  %\setlength{\belowcaptionskip}{2pt}
  %\vspace*{-1.8cm}
  %\caption{Accuracy}
  %\label{fig:acc}
%\end{subfigure}%
%\hspace*{-13em}
%\begin{subfigure}{0.5\textwidth}
%\rule{\linewidth}{0.5\linewidth}
 % \centering
 % \includegraphics[width=1.1\linewidth]{figures/time_performance.pdf}
 % \vspace*{-1.8cm}
 % \caption{Time Performance}
 % \label{fig:timePerform}
%\end{subfigure}

%\begin{subfigure}{0.5\textwidth}
%\rule{\linewidth}{0.5\linewidth}
 % \centering
  %\includegraphics[width=1.1\linewidth]{figures/robustness.pdf}
  %\vspace*{-1.8cm}
 % \caption{Robustness}
 % \label{fig:robustness}
%\end{subfigure}

%\caption{Performance Metrics}
%  \label{fig:perf-metrics}
%\end{figure*}
\begin{figure}
    \centering
    \includegraphics[width=1.1\columnwidth]{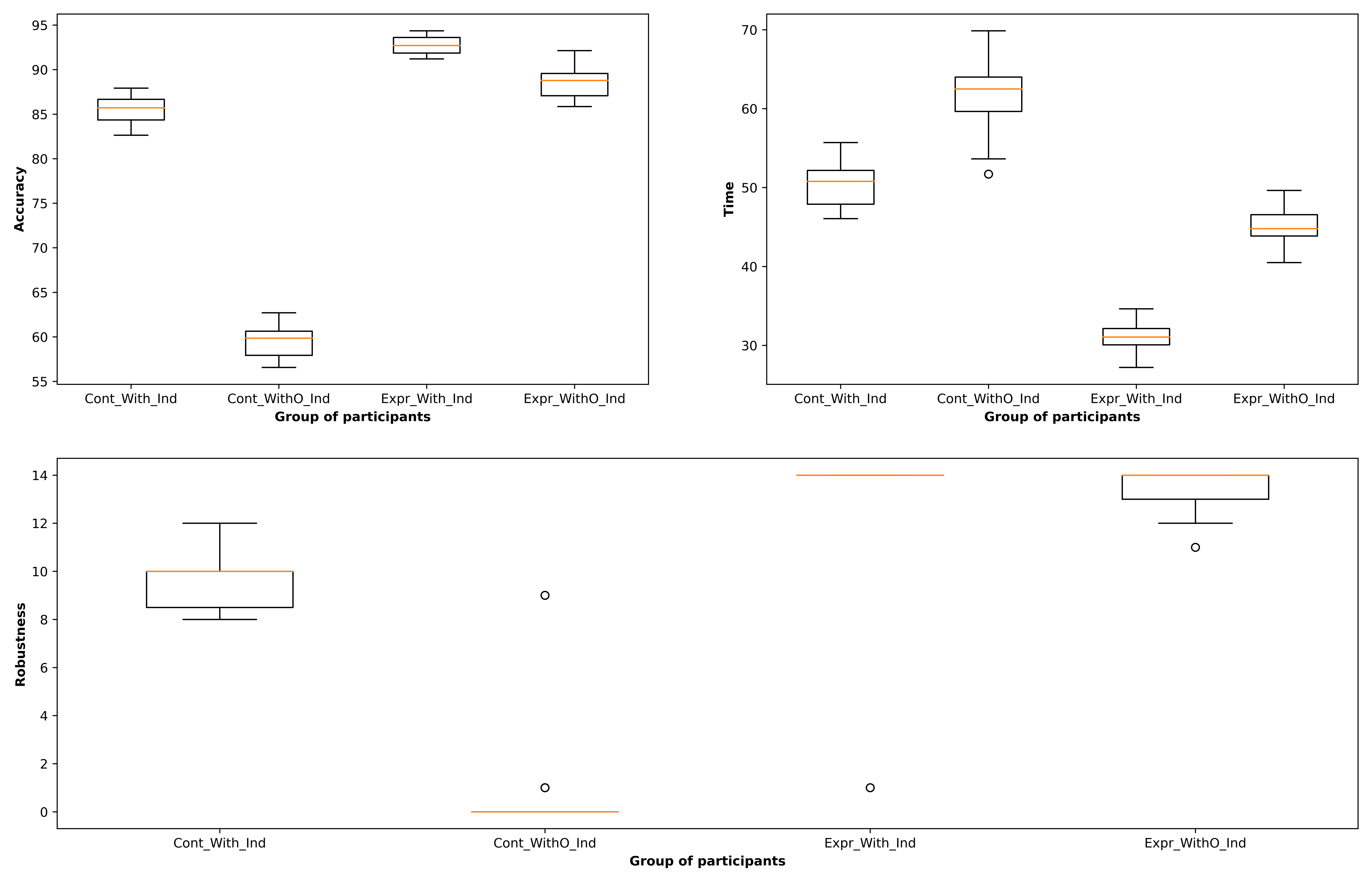}
    \caption{Performance Metrics}
    \label{fig:perf-metrics}
\end{figure}
\subsection{Procedure}
As part of our experiments, we distributed various architectural diagrams to the controlled group  and the experimental group. Each of the participants was given 28 architectural diagrams, two from each of the 14  categories. We asked them to identify the significant architectural components, patterns, and overall system properties (quality attributes and tactics) perceived by them after analyzing the architectural diagrams without any additional documents. The participants were free to use any other information source available to them with appropriate citations. We manually evaluated all the assignment responses and scored them based on the following parameters for each architectural diagram to measure the accuracy and explainability: 
\begin{enumerate}
    \item The number of components and their characteristics correctly identified.
    \item The number of connections and their characteristics correctly identified.
    \item The number of design patterns correctly identified with justification.
    \item The number of quality attributes correctly identified with justification. 
    \item The number of architectural tactics correctly identified with justification. 
\end{enumerate}
We also asked each participant to share the time spent on the evaluation of each architectural image  to measure the time performance.

\subsection{Results and Discussion}
Each of the box plot graphs shown in Fig. \ref{fig:perf-metrics} represents the distribution of actual values at a minimum value, 25 percentile, 75 percentile, and maximum values, respectively. The Groups of participants are annotated as: 
\begin{itemize}
    \item Controlled group with Industrial Experience: Cont\_With\_Ind
    \item Controlled group without Industrial Experience: Cont\_WithO\_Ind
    \item Experimental group with Industrial Experience: Exp\_With\_Ind
    \item Experimental group without Industrial Experience:
    Exp\_WithO\_Ind
\end{itemize}

\subsubsection{Accuracy and Explainability}
The responses of each candidate and inference outcome of the automated model were evaluated by teaching assistants manually for their accuracy. 

\textbf{Salient Observations:}
\begin{enumerate}
    \item The Controlled group without Industrial Experience, i.e., the group without Industrial Experience and without the help of our framework performs the worst, while the Experimental Group With Industrial Experience performs the best achieving an highest accuracy of 94.37\%.
    \item As the accuracy of Exp\_With\_Ind group is more than the accuracy of Exp\_WithO\_Ind, and a similar trend is observed in case of controlled groups, it can be concluded that the accuracy of the group improves with the help of our framework. 
    \item The improvement in accuracy as observed in the category of the candidates Without Industrial experience, viz., Cont\_WithO\_Ind to Exp\_WithO\_Ind is huge (48.9\% on average). Therefore, our framework contributes significantly in improving the design decision process of the novice software architects.
    \item The accuracy improvement in case of candidates with prior industrial experience, viz., Cont\_With\_Ind to Exp\_With\_Ind is 8.42\%, and thus is beneficial in improving the decision process of industrial experts as well.
\end{enumerate}

\textbf{Inference:} It is evident that the usage of our proposed model to evaluate the architectural designs reduces the variation in the results and, at the same time, increases the overall correctness of the responses. The explainability score indicates that the experimental group participants are performing better than the controlled group participants.

\subsubsection{Time Performance}

%\hspace{-2cm}
The results depicted in the time boxplots of the figure validate that the experimental group performed far better than the controlled group and took lesser time in evaluation the architectural designs (47.8\% improvement on average). Further, the use of the proposed approach helped speed up the evaluation process (37.82\% for novice architects and 62.2\% for experienced architects). Hence, it shows that using the proposed approach will help novice and experienced candidates to evaluate the architectural designs in lesser time as compared to when performed withour our framework's support. Also, the use of our framework leads to results with limited variation in the responses in terms of time, efforts, and accuracy.

\textbf{Critical Reasoning:} The time improvement is more in case of experienced architects, which might imply that using the proposed framework experts were able to evaluate the design diagrams in a considerable lesser time. The reduced variation the results leads to an increase in the certainty in effort and time estimation for the architectural evaluation tasks, and help in better resource and cost planning. 

\subsubsection{Robustness}

 We included the robustness parameter to avoid bias due to candidates' prior experience in the control and experimental groups. Candidates might be good at evaluating the architectural patterns that they have worked with in the past. This parameter analyzed whether a candidate could answer with higher accuracy across the different categories of architectural design images. To measure this parameter, we aggregated each candidate's score for each category and calculated the percentage of categories for which they have scored more than 80\%. As shown in the robustness boxplots, the experimental group shows a better hold on robustness than the controlled group (158.6\% average improvement). The largest improvement was observed from the Cont\_WithO\_Ind group to Exp\_WithO\_Ind of 1953.85\%, with the respective average robustness scores as 0.65 vs. 13.35.  

\textbf{Inference:} A significant improvement in robustness scores for the novice architects signifies that the framework helps them to understand and evaluate the design diagrams from different architectural patterns effectively. It also in a way validates that the framework is not biased to support the decision making process of any specific architectural pattern type.

\section{Threats to validity}

Several threats affect the validity of this work. Firstly, the scope of our work is limited only to the considered set of architectural patterns. Our framework cannot detect the patterns not considered currently and thus does not comment about their QAs or tactics. However, this can be overcome by extending our study for the left-out architectural patterns by appending the dataset with appropriate images. 

Secondly, the images have been acquired from the web, which calls for many challenges, such as size or resolution variation, illumination variation, blurriness, noise, and color in the images, affecting the overall process;  because feature extraction is primarily dependent on the quality of the images. Therefore, manual filtering for removing the images not projecting the considered patterns and low quality is performed. 

Further, the pre-processing of acquired images is done in order to enhance them for better feature extraction. Another threat could be the use of OCR in the mapping procedure. The images can have labels written in various font styles and sizes, or it could be the case that some images may contain handwritten labels. Many independent variables were used in these models, which could affect the accuracy of predictions. However, we used step-wise techniques to identify the optimum number of variables for the models and tested multi-col-linearity models.  These risks were partially mitigated using a classic N-cross-fold validation to evaluate the models and demonstrated that they were generalizable. 

Lastly, the experiments conducted with the controlled group and experimental group may have some limitations. We asked candidates to follow the guidelines as much as possible to ensure that equivalence among them in terms of resources available to them. Bigger group size should have been better to reduce noisy and biased responses from the results. We had to discard the three submissions from both the groups due to the lack of enough data for estimating time performance and robustness.  

\section{Conclusion and Future Work}

An automated framework to evaluate a software architecture with image processing, and inference from the QA knowledge-base is quite helpful. It reduces the variation among responses and increases the accuracy across the multiple categories of architectural patterns. Further, the effectiveness of architecture image matching could be improved by exploring graph-based techniques on a larger image dataset that can connect vision components and determine combinatorial similarity. Possible improvements using OCR methods could be explored for better handling of design artifacts and handwritten annotations. These improvements will help increase the explainability of our framework and improve the overall model's robustness. Developing the framework in the form of a full-fledged automated application for community use is a part of our future work. We are currently working on the system prototype and are planning to add more architectural knowledge artifacts to increase the effectiveness of this approach and provide more advanced recommendations.
\bibliographystyle{splncs04}
\bibliography{bibFile}

\begin{thebibliography}{10}
\providecommand{\url}[1]{\texttt{#1}}
\providecommand{\urlprefix}{URL }
\providecommand{\doi}[1]{https://doi.org/#1}

\bibitem{abowd1997recommended}
Abowd, G., Bass, L., Clements, P., Kazman, R., Northrop, L.: Recommended best
  industrial practice for software architecture evaluation. Tech. rep.,
  Carnegie-Mellon Univ Pittsburgh Pa Software Engineering Inst (1997)

\bibitem{bachmann2003deriving}
Bachmann, F., Bass, L., Klein, M.: Deriving architectural tactics: A step
  toward methodical architectural design. Tech. rep., Carnegie-Mellon Univ
  Pittsburgh Pa Software Engineering Inst (2003)

\bibitem{bachmann2007modifiability}
Bachmann, F., Bass, L., Nord, R.: Modifiability tactics. Tech. rep.,
  CARNEGIE-MELLON UNIV PITTSBURGH PA SOFTWARE ENGINEERING INST (2007)

\bibitem{bass2003software}
Bass, L., Clements, P., Kazman, R.: Software architecture in practice.
  Addison-Wesley Professional (2003)

\bibitem{bengtsson1998scenario}
Bengtsson, P., Bosch, J.: Scenario-based software architecture reengineering.
  In: Proceedings. Fifth International Conference on Software Reuse (Cat. No.
  98TB100203). pp. 308--317. IEEE (1998)

\bibitem{bi2018architecture}
Bi, T., Liang, P., Tang, A.: Architecture patterns, quality attributes, and
  design contexts: How developers design with them. In: 2018 25th Asia-Pacific
  Software Engineering Conference (APSEC). pp. 49--58. IEEE (2018)

\bibitem{bochinski2017high}
Bochinski, E., Eiselein, V., Sikora, T.: High-speed tracking-by-detection
  without using image information. In: 2017 14th IEEE International Conference
  on Advanced Video and Signal Based Surveillance (AVSS). pp.~1--6. IEEE (2017)

\bibitem{bosch2000design}
Bosch, J.: Design and use of software architectures: adopting and evolving a
  product-line approach. Pearson Education (2000)

\bibitem{capilla2020teaching}
Capilla, R., Zimmermann, O., Carrillo, C., Astudillo, H.: Teaching students
  software architecture decision making. In: European Conference on Software
  Architecture. pp. 231--246. Springer (2020)

\bibitem{chung2012non}
Chung, L., Nixon, B.A., Yu, E., Mylopoulos, J.: Non-functional requirements in
  software engineering, vol.~5. Springer Science \& Business Media (2012)

\bibitem{ieee1990ieee}
Committee, I.S.C., et~al.: Ieee standard glossary of software engineering
  terminology (ieee std 610.12-1990). los alamitos. CA: IEEE Computer Society
  \textbf{169} (1990)

\bibitem{dobrica2002survey}
Dobrica, L., Niemela, E.: A survey on software architecture analysis methods.
  IEEE Transactions on software Engineering  \textbf{28}(7),  638--653 (2002)

\bibitem{fielding2000architectural}
Fielding, R.T., Taylor, R.N.: Architectural styles and the design of
  network-based software architectures, vol.~7. University of California,
  Irvine Irvine (2000)

\bibitem{gonzalez2013defining}
Gonz{\'a}lez-Huerta, J., Insfr{\'a}n, E., Abrah{\~a}o, S.: Defining and
  validating a multimodel approach for product architecture derivation and
  improvement. In: International Conference on Model Driven Engineering
  Languages and Systems. pp. 388--404. Springer (2013)

\bibitem{jin2021image}
Jin, Y., Mishkin, D., Mishchuk, A., Matas, J., Fua, P., Yi, K.M., Trulls, E.:
  Image matching across wide baselines: From paper to practice. International
  Journal of Computer Vision  \textbf{129}(2),  517--547 (2021)

\bibitem{kalra2018towards}
Kalra, S., Prabhakar, T.: Towards dynamic tenant management for microservice
  based multi-tenant saas applications. In: Proceedings of the 11th Innovations
  in Software Engineering Conference. pp.~1--5 (2018)

\bibitem{kapur2019towards}
Kapur, R., Sodhi, B.: Towards a knowledge warehouse and expert system for the
  automation of sdlc tasks. In: 2019 IEEE/ACM International Conference on
  Software and System Processes (ICSSP). pp.~5--8. IEEE (2019)

\bibitem{kapur2020defect}
Kapur, R., Sodhi, B.: A defect estimator for source code: Linking defect
  reports with programming constructs usage metrics. ACM Transactions on
  Software Engineering and Methodology (TOSEM)  \textbf{29}(2),  1--35 (2020)

\bibitem{kapur2021using}
Kapur, R., Sodhi, B., Rao, P.U., Sharma, S.: Using paragraph vectors to improve
  our existing code review assisting tool-cruso. In: 14th Innovations in
  Software Engineering Conference (formerly known as India Software Engineering
  Conference). pp. 1--11 (2021)

\bibitem{kazman1994saam}
Kazman, R., Bass, L., Abowd, G., Webb, M.: Saam: A method for analyzing the
  properties of software architectures. In: Proceedings of 16th International
  Conference on Software Engineering. pp. 81--90. IEEE (1994)

\bibitem{kazman1998architecture}
Kazman, R., Klein, M., Barbacci, M., Longstaff, T., Lipson, H., Carriere, J.:
  The architecture tradeoff analysis method. In: Proceedings. Fourth IEEE
  International Conference on Engineering of Complex Computer Systems (Cat. No.
  98EX193). pp. 68--78. IEEE (1998)

\bibitem{kim2009quality}
Kim, S., Kim, D.K., Lu, L., Park, S.: Quality-driven architecture development
  using architectural tactics. Journal of Systems and Software  \textbf{82}(8),
   1211--1231 (2009)

\bibitem{kruchten19954+}
Kruchten, P.B.: The 4+ 1 view model of architecture. IEEE software
  \textbf{12}(6),  42--50 (1995)

\bibitem{lassing1999software}
Lassing, N., Rijsenbrij, D., van Vliet, H.: On software architecture analysis
  of flexibility, complexity of changes: Size isn't everything. In: Proceedings
  of the second Nordic Software Architecture Workshop (NOSA'99) Ronneby,
  Sweden, 1999 (1999)

\bibitem{li2020understanding}
Li, S., Zhang, H., Jia, Z., Zhong, C., Zhang, C., Shan, Z., Shen, J., Babar,
  M.A.: Understanding and addressing quality attributes of microservices
  architecture: A systematic literature review. Information and Software
  Technology p. 106449 (2020)

\bibitem{lowe1999object}
Lowe, D.G.: Object recognition from local scale-invariant features. In:
  Proceedings of the seventh IEEE international conference on computer vision.
  vol.~2, pp. 1150--1157. Ieee (1999)

\bibitem{lung1997approach}
Lung, C.H., Bot, S., Kalaichelvan, K., Kazman, R.: An approach to software
  architecture analysis for evolution and reusability. In: Proceedings of the
  1997 conference of the Centre for Advanced Studies on Collaborative research.
  p.~15 (1997)

\bibitem{maggiori2014towards}
Maggiori, E., Gervasoni, L., Ant{\'u}nez, M., Rago, A., D{\'\i}az~Pace, J.A.:
  Towards recovering architectural information from images of architectural
  diagrams. Jornadas Argentinas de Inform{\'a}tica e Investigaci{\'o}n
  Operativa (JAIIO)(Argentina)  \textbf{43} (2014)

\bibitem{mark2015software}
Mark, R.: Software architecture patterns-understanding common architecture
  patterns and when to use them (2015)

\bibitem{molter1999integrating}
Molter, G.: Integrating saam in domain-centric and reuse-based development
  processes. In: Proceedings of the 2nd Nordic Workshop on Software
  Architecture, Ronneby. pp. 1--10 (1999)

\bibitem{osses2018exploration}
Osses, F., M{\'a}rquez, G., Astudillo, H.: Exploration of academic and
  industrial evidence about architectural tactics and patterns in
  microservices. In: Proceedings of the 40th International Conference on
  Software Engineering: Companion Proceeedings. pp. 256--257 (2018)

\bibitem{7332519}
Panichella, S.: Supporting newcomers in software development projects. In: 2015
  IEEE International Conference on Software Maintenance and Evolution (ICSME).
  pp. 586--589 (2015). \doi{10.1109/ICSM.2015.7332519}

\bibitem{pinto2019training}
Pinto, G., Ferreira, C., Souza, C., Steinmacher, I., Meirelles, P.: Training
  software engineers using open-source software: the students' perspective. In:
  2019 IEEE/ACM 41st International Conference on Software Engineering: Software
  Engineering Education and Training (ICSE-SEET). pp. 147--157. IEEE (2019)

\bibitem{pinto2017training}
Pinto, G.H.L., Figueira~Filho, F., Steinmacher, I., Gerosa, M.A.: Training
  software engineers using open-source software: the professors' perspective.
  In: 2017 IEEE 30th Conference on Software Engineering Education and Training
  (CSEE\&T). pp. 117--121. IEEE (2017)

\bibitem{reussner2003reliability}
Reussner, R.H., Schmidt, H.W., Poernomo, I.H.: Reliability prediction for
  component-based software architectures. Journal of systems and software
  \textbf{66}(3),  241--252 (2003)

\bibitem{scott2009realizing}
Scott, J., Kazman, R.: Realizing and refining architectural tactics:
  Availability. Tech. rep., CARNEGIE-MELLON UNIV PITTSBURGH PA SOFTWARE
  ENGINEERING INST (2009)

\bibitem{selby1995interconnectivity}
Selby, R.W., Reimer, R.M.: Interconnectivity analysis for large software
  systems. In: Proceedings of the California Software Symposium. pp. 3--17
  (1995)

\bibitem{trowbridge2004integration}
Trowbridge, D.: Integration patterns. Microsoft Press (2004)

\bibitem{ye2017robust}
Ye, Y., Shen, L., Hao, M., Wang, J., Xu, Z.: Robust optical-to-sar image
  matching based on shape properties. IEEE Geoscience and Remote Sensing
  Letters  \textbf{14}(4),  564--568 (2017)

\end{thebibliography}
\end{document}